\documentclass[pre,aps,showpacs,byrevtex,amsmath,amssymb,twocolumn]{revtex4}
\usepackage{graphicx}
\usepackage{amsmath}
\usepackage{amsfonts}
\usepackage{amssymb}

\begin{document}

\title{Kinetics of non-ionic surfactant adsorption at a fluid-fluid interface from a micellar solution}
\author{Herv\'{e} Mohrbach  }
\affiliation{Instituut-Lorentz, Universiteit Leiden, Postbus $9506$, $2300$ RA Leiden, The netherlands}

\begin{abstract}
The kinetics of non-ionic surfactant adsorption at a fluid-fluid interface from a micellar solution is considered theoretically.
Our model takes into account the effect of micelle relaxation on the diffusion of the free surfactant molecules. It is shown that
non-ionic surfactants undergo either a diffusion or a kinetically limited adsorption according to the characteristic relaxation time of the micelles. This gives a new interpretation for the observed dynamical surface tension of micellar solutions.

\end{abstract}

\maketitle Since the pioneering work of Ward and Torday \cite{wt}
many theories have been suggested to describe the kinetics of
non-ionic surfactant adsorption at a fluid-fluid interface
\cite{theories}. More recently Andelman and Diamant \cite{ad}
presented a free energy formulation of the kinetics of
surfactant molecule adsorption which describes the diffusive transport
inside the solution and at the same time the adsorption kinetics
at the interface itself. Their approach which was successfully
applied to the kinetics of non-ionic surfactant \cite{ad1,ad2,ad3}
allows them to treat also the case of ionic surfactant adsorption
\cite{ad4}. These previous theories predict the adsorption kinetics
for dilute systems only, without aggregation in the bulk. Numerous
dynamic measurements of the surface tension show that above the critical
micellar concentration ($cmc$),
micelles, although having a negligible surface activity, increase
the adsorption rate of free surfactant molecules (\cite{Noskov}
and references therein). A widely accepted physical picture which
explains this effect is the following: above the critical micellar
concentration surfactant molecules associate into micelles which
are in equilibrium with free surfactants whose concentration $\phi
_{b}$ is approximatively equal to the $cmc$. Adding more monomers in 
the solution leads to the formation of new micelles. In the presence 
of an interface it
is usually assumed that only free surfactants adsorb, because the
adsorption energy of the micelles is too high \cite{mukerjee}.
Therefore at the early stages, molecules diffuse and adsorb at the
interface. Then locally there is a depletion layer where the
concentration of molecules falls below the $cmc$. The micelles in
this region become unstable and relax, releasing surfactant
molecules in order to redress the equilibrium, but these released
molecules can themselves adsorb. Thus micelles behave like
reservoirs of monomers that try to maintain the volume fraction of
surfactant molecules at the $cmc$. At the same time, the micelles
will also diffuse as their concentration become non-uniform. In
most of the previous theoretical treatments, this diffusion of
micelles is neglected and we will make here the same
approximation. Therefore the diffusion equation of the monomers
must contain terms that account for the presence of the micelles.
Previous theories proposed for this purpose
\cite{Noskov,Joos,Petkova} assume a diffusion-limited adsorption
which implies that the interface is in equilibrium at all time
with the adjacent solution. But some experiments show a weak
concentration dependence of the dynamic surface tension
\cite{Fainerman} indicating that the adsorption might be
kinetically-limited.

In the current letter we extend the theory of Andelman and Diamant \cite{ad} to include
the kinetics of surfactant adsorption from a micellar solution. This will allow us to
determine under which condition the adsorption is diffusion or kinetically limited.

Since the extent of the influence of the micelles on the adsorption kinetics depends on the rates of micelle relaxation we must first consider the model of Aniansson and Wall \cite{aniansson,aniansson2,aniansson3} which distinguishes a slow and a fast process of
relaxation (see also \cite{Noskov} and \cite{joanny}). In this model the creation rate of free surfactants can be written
\begin{equation}
\frac{d\phi }{dt}=-\frac{\left( \phi(t) -\phi _{b}\right) }{\tau_{1,2}} \label{creationrate}
\end{equation}
where $\phi(t)$ is the concentration of free monomers a time $t$ and $\phi _{b}$ is their equilibrium concentration approximatively equal to the $cmc$. The relaxation 
times $\tau_{1}$ and $ \tau_{2}$ correspond
to the fast and the slow process respectively and depend upon both 
the average extraction time of a single chain and the micellar size distribution. 
During the fast step of the relaxation process ($\tau_{1}\approx 10^{-6}$s for
common non-ionic surfactant) each micelle releases only a few
surfactant molecules in the solution so that the total amount of
micelles remains constant. Further evolution of the system
consists of a slow change of the total number of micelles by two
mechanisms: the release of molecule by molecule from the micelle and
the fission process \cite{Lessner} where the micelle splits into
smaller pieces. For conventional neutral surfactant the fast
process is beyond the time interval experimentally accessible.
Besides, for small concentrations one should expect of very small
effect of the fast process on the kinetics of adsorption as it
consumes only a small amount of monomers compared to the slow
process for which $\tau_{2}\approx 10^{-3}$s \cite{Noskov}. Then
in the current work we will consider only the slow process.

To be concrete we consider an infinite volume with a flat
interface lying in the plane $x=0$ and treat the bulk solution and
the interface as two coupled interdependent sub-systems. The
volume fraction of free surfactants can be considered as dilute since
$\phi _{b}\approx cmc$ is usually small, whereas the volume fraction at the
interface $\phi_0$ may become large. The system is in contact with
a reservoir where the total volume fraction of surfactant is
fixed. For a sub-micellar solution, 
the volume fraction $\phi$ being a conserved quantity, the kinetic equation for a conserved order parameter can be employed \cite{ad,ad1,ad2,ad3,ad4}. 
For a micellar solution, we model micelles as sources of surfactant molecules which are present anywhere in the solution and thus violate mass conservation (the bulk volume fraction is now a non-conserved order parameter). We thus assume a diffusion-relaxation equation for the surfactant molecules in the bulk \cite{Joos,Petkova}
\begin{equation}
\frac{\partial \phi }{\partial t}=D\frac{\partial ^2\phi}{\partial x^2}-\frac{
\left( \phi -\phi _{b}\right) }{\tau }  \label{diff}
\end{equation}
where $D$ is the free surfactant diffusivity. The source term accounts for the micelle relaxation on the adsorption kinetics of the surfactant molecules. The characteristic time scale $\tau=\tau_2$ of relaxation determines the extent of the influence of
micelles on the adsorption kinetics.

Close to the interface we consider as in \cite{ad} a discretized current densities and write the dynamic surface coverage as ($j_0=0$ in the absence of evaporation) 
\begin{equation}
\frac{\partial \phi_0 }{\partial t}=-\frac{j_1}{a}  \label{dphio}
\end{equation}
where $\phi_0$ is the interfacial volume fraction, $a$ is the surfactant molecular dimension and $j_1$ the current density driving the surfactant molecules between the sub-surface layer $x=a$ to the interface $x=0$.
From the conservation condition at the sub-surface layer we get
\begin{equation}
\frac{\partial \phi _{1}}{\partial t}=\frac{j_{1}-j_{2}}{a}=D\left. \frac{
\partial \phi }{\partial x}\right| _{a}-\frac{\partial \phi _{0}}{\partial t}
.  \label{dphi1}
\end{equation}
where $\phi _{1}$ is the volume fraction in the sub-surface layer.

Assuming an initial uniform state $\phi_{b}$ and applying Laplace transform
method to the system of equations (\ref{diff}) and (\ref{dphi1}), an
expression is obtained relating the surface density to the sub-surface volume
fraction:
\begin{eqnarray}
\phi _{0}(t) &=&\frac{-\sqrt{D}}{a\sqrt{\pi }}\left\{ \int_{0}^{t}dt'
\left( \phi _{1}(t')-\phi _{b}\right) \left[ \frac{e^{-\gamma \left(
t-t' \right) }}{\sqrt{\left( t-t' \right) }} \right. \right.  \nonumber\\
&& \left.\left. +\gamma \int_{0}^{t-t'
}dt_{1}\frac{e^{-\gamma t _{1}}} {\sqrt{t_{1}}}\right] \right\}-\phi _{1}\left( t\right) +2\phi _{b}  \label{equ}
\end{eqnarray}
where $\gamma =1/\tau$.
\bigskip

The current density between the sub-surface layer and the interface must be defined in a discret manner since the two surfaces are separated by a distance of the size of the monomeres. Therefore as in \cite{ad} we write
\begin{equation}
j_1=\frac{D_0}{T}(\mu _{1}-\mu _{0})/a
\end{equation}
with $D_0$ the diffusivity at the interface. The chemical potential at the sub-surface layer $x=a$ is \begin{equation}
\mu _{1}=T\ln{\phi_1}
\end{equation}
as the surfactant solution can be considered as dilute and ideal. At the interface itself this is no more the case as $\phi_0$ may become large. The chemical potential $\mu _{0}$ can be determined by considering as in \cite{ad} the free energy per unit area of the surfactant monomers at the interface:
\begin{eqnarray}
f_{0}\left( \phi _{0}\right) &=& \frac{1}{a^{2}}\left\{ T\left[ \phi _{0}\ln
\phi _{0}+\left( 1-\phi _{0}\right) \ln \left( 1-\phi _{0}\right) \right] \right.  \nonumber\\
&& \left. -\alpha \phi _{0}-\frac{\beta }{2}\phi _{0}^{2}-\mu _{1}\left( t\right) \phi
_{0}\right\}   \label{deltaf0}
\end{eqnarray}
The term in the square bracket is the entropy of mixing and the
parameters $\alpha $ and $\beta $ are respectively the affinity of the
molecules for the interface and the lateral interaction between two adjacent
surfactant molecules. The chemical potential $\mu _{1}$ accounts
for the contact of the interface to the adjacent surface ($x=a$). The excess in chemical potential at the interface result from the variation of $f_0$ with respect to $\phi _{0}$ which yields to the following chemical potential
\begin{equation}
\mu _{0}=T\ln \frac{\phi _{0}}{1-\phi _{0}}-\alpha -\beta \phi _{0} \label{mu}
\end{equation}
Then from eq.~\ref{dphio} we get the equation of Andelman and Diamant governing the adsorption kinetics at the interface 
\begin{equation}
\frac{\partial \phi _{0}}{\partial t}=\frac{-j_{1}}{a}=\frac{D_0}{a^{2}}\phi
_{1}\left[ \ln \frac{\phi _{1}\left( 1-\phi _{0}\right) }{\phi _{0}}+\frac{%
\alpha +\beta \phi _{0}}{T}\right]  \label{cinetique}
\end{equation}
The time-dependent surface coverage in a micellar phase is therefore solution of the system of two coupled equations (\ref{equ}) and (\ref{cinetique}). Whereas equation (\ref{equ}) describes the kinetics taking place inside the bulk solution and generalizes the results of \cite{ad} to a micellar phase, equation (\ref{cinetique}) which is valid for a submicellar as well as for a micellar solution, describes the adsorption kinetics at the interface.

\smallskip

In the micellar phase, we have three time scales, $\tau _{d}$, $\tau _{k}$ and $\tau $, that characterize respectively the diffusion in the bulk, at the interface and the micelle relaxation. These time scales determine limiting cases that we now propose to explore. The time scale $\tau _{k}$ of the kinetics at the interface has been determined in \cite{ad} by looking at the asymptotic behavior of equation (\ref{cinetique}) which yields
\begin{equation}
\phi _{0,eq}-\phi _{0}\left( t\right) \sim e^{-t/\tau _{k}}  \label{tauk}
\end{equation}
where $\phi _{0,eq}$ is the equilibrium surface coverage and
\begin{equation}
\tau _{k} \approx  \frac{a^{2}\phi _{0,eq}^{2}}{D_0\phi _{b}^{2}}e^{-\left( \alpha
+\beta \phi _{0,eq}\right) /T}. \label{ts}
\end{equation}
When $\tau _{k}$ is the dominant time scale, the kinetic process at the interface is the slower one. In this case the concentration inside the bulk is assumed to be constant ($\phi_1=\phi_b$) and the dynamic surface is given by eq. \ref{cinetique} only. The adsorption process is then kinetically limited (KLA process).

In order to estimate the time scale of diffusion $\tau _{d}$, we first consider the adsorption kinetics of dilute solution below the $cmc$ which can be rederived by considering equation (\ref{equ}) with $\gamma=0$. This leads to the result of Andelman and Diamant \cite{ad} similar to the one obtained previously by Ward and Torday \cite{wt}
\begin{equation}
\phi _{0}\left( t\right) =\frac{1}{a}\sqrt{\frac{D}{\pi }}\left[ 2\phi_b \sqrt{
t}-\int_{0}^{t}\frac{\phi _{1}\left( t' \right) }{\sqrt{t-t' }}dt'
\right] +2\phi_b -\phi _{1}\left( t\right)   \label{wt}
\end{equation}
where $\phi_b$ is now smaller than the $cmc$.
The asymptotic limit of eq.~\ref{wt}
\begin{equation}
\phi _{b}-\phi _{1} \sim \frac{a\phi _{0,eq}}{\sqrt{\pi Dt}} \label{gama1}
\end{equation}
allows us to identify the characteristic time scale of diffusion \cite{ad}
\begin{equation}
\tau_d =(\frac{\phi _{0,eq}}{\phi _{b}})^2\frac{a^{2}}{D} \label{gama}
\end{equation}
The comparison of eqs.~\ref{ts} and \ref{gama1} leads to $\tau_d>\tau _{k}$ (since $D$ and $D_0$ are usually of the same order). This result implies that common non-ionic surfactants in a non micellar solution exhibit diffusion limited adsorption (DLA) \cite{ad}. It means that the equilibration process inside the solution is slower than the one at the interface. Remarkably, for a solution of ionic surfactant without added salt, Andelman and Diamant predicted  a kinetically limited adsorption in agrement with experiments. This effect is due to the strong electrostatic interactions which drastically accelerate the kinetics in the bulk. As shown in the following, a strong acceleration of the kinetics is also induced by the presence of micelles, an effect that may make the surface kinetics become the limiting process for adsorption (KLA process).

\smallskip
We now return to a micellar solution and from eq.~ \ref{equ} we observe two different regimes of the adsorption process:
\smallskip

$(i)$ for time $t<<\tau,$ eq.~\ref{equ} reduces to eq.~\ref{wt} since the micelle relaxation mechanism is not yet relevant. In this regime the process is always of DLA type \cite{ad}.

$(ii)$ in the micelle-limited regime $t\gg \tau$, the surface coverage eq~(\ref{equ}) can be approximated by
\begin{equation}
\phi _{0}\left( t\right) \approx \frac{\sqrt{D\gamma }}{a}\left[ \phi
_{b}t-\int_{0}^{t}\phi _{1}\left( \tau \right) d\tau \right] -\phi
_{1}\left( t\right) +2\phi _{b}.  \label{phiom}
\end{equation}
This expression predicts a strong increase of the diffusion kinetics in the bulk as compared to eq. \ref{wt}. This behavior of the adsorption kinetic (proportional to $\sqrt{t}$ at short times and proportional to $t$ at long times) is frequently observed in experiments \cite{Joos}. Note that the solution eq.~\ref{phiom} could also be deduced by neglecting the time derivative of the left-hand side of equation (\ref{diff}); the surfactant molecule profile being then quasi-steady inside the solution.

Let us examine the characteristic time scale associated with the kinetic equation (\ref{phiom}). For a DLA process, the asymptotic behavior of eq.~\ref{phiom} leads to the following asymptotic time dependence
\begin{equation}
\phi _{b}-\phi _{1}\left( t\right) \sim
\frac{a\phi _{0,eq}e^{-t/\tau}}{\sqrt{\pi Dt}}.  \label{gama2}
\end{equation}
Remarkably the presence of the exponential term dramatically shortens the time scale of
diffusion inside the solution. We can then conclude that the micelles play a major
role in the adsorption process by drastically accelerating the kinetics
inside the solution. That the kinetics inside the bulk is increased does not mean that it is necessarily faster than the kinetics at the interface itself, i.e., that the adsorption is kinetically-limited.

To solve this issue we must consider the following two situations:

(i) $\tau>\tau_k$. In this case we have a DLA process. We can assume that the interface is in equilibrium at all time with the sub-surface layer and $\phi_1$ is related to $\phi_0$ by an isotherm given by the condition $\mu_1=\mu_0$ \cite{ad}, which yields
\begin{equation}
\phi _{0}(t)=\frac{\phi _{1}(t)}{\phi _{1}(t)+e^{-\left( \alpha +\beta \phi
_{0}(t)\right) /T}} \label{isotherme}
\end{equation}
The combination of this equation with eq.~\ref{equ} amounts to solve the problem of the adsorption kinetic in a micellar phase for a DLA process. For time $\tau_k<t<\tau$ the micelle relaxation mechanism is not yet relevant and equation (\ref{equ}) reduces to eq (\ref{wt}).
The kinetics is the one of a submicellar solution. For time $\tau<t$, eq.~\ref{equ} is approximated by eq.~\ref{phiom} as the micelle relaxation mechanism becomes relevant and the kinetics in the bulk solution is strongly accelerated. Nevertheless it is slower than the kinetics at the interface as $\tau>\tau_k$. Therefore adsorption kinetics of a micellar solution of non-ionic surfactants which undergo a diffusion limited adsorption show two different regimes, in accord with experimental findings \cite{Joos}; in particular, experimental curves representing the surface coverage as a function of time show a dependence in the total concentration of surfactant \cite{Joos,Noskov}. Our model can explain this dependence as the surface coverage eq.~\ref{phiom} depends on the time scale $\tau$ which itself varies with the total concentration of surfactant \cite{Noskov,aniansson,aniansson2,aniansson3,joanny,Lessner}.
As shown in \cite{ad}, the dynamical surface tension in a DLA process can be approximated by the interfacial free energy $\Delta \gamma \left( t\right)\approx f_{0}\left( \phi_{0}\left( t\right) \right)$ which yields
\begin{equation}
\Delta \gamma(t)\approx \frac{1}{a^{2}}\left[ T\ln \left( 1-\phi_{0}\right) +\frac{\beta }{2}\phi_{0}^{2}\right]
\end{equation}
where $\phi_0$ is given by the solution of the system of equations (\ref{equ}) and (\ref{isotherme}).

(ii) $\tau<\tau_k$. In the micelle-limited regime ($\tau<t$) we have a KLA process, where the solution is assumed to be at all time in equilibrium with the bulk reservoir. Then $\phi(x)= \phi_b$ at all time and $\phi_0$ changes according to eq.~\ref{cinetique}, that we write now
\begin{equation}
\frac{\partial \phi _{0}}{\partial t}=\frac{D}{a^{2}}\phi _{b}\left[ \ln
\frac{\phi _{b}\left( 1-\phi _{0}\right) }{\phi _{0}}+\frac{\alpha +\beta
\phi _{0}}{T}\right] \label{cinetique2}
\end{equation}
The time dependent surface coverage can then be obtained by a single integration. Importantly the surface coverage as a function of time is $\tau$ independant. Actually eq.~\ref{cinetique2} depends only on $\phi_b$ (which is almost equal to the $cmc$) whose variation with the total concentration is very small. In this context, the dynamic surface tension is simply given by the interfacial free energy $\Delta \gamma \left( t\right)=f_{0}\left( \phi_{0}\left( t\right) \right)$ which yields
\begin{eqnarray}
\Delta \gamma \left( t\right) &=&\frac{1}{a^{2}}\left\{ T\left[ \phi _{0}\ln \phi _{0}+\left( 1-\phi_{0}\right) \ln \left( 1-\phi_{0}\right) \right] \right.  \nonumber \\
&&\left. -\alpha \phi_{0}-\frac{\beta }{2}\phi_{0}^{2}-T\phi_{0}\ln\phi_{b} \right\}.
\end{eqnarray}
The dynamical surface tension as a function of time is therefore independant of the total surfactant concentration. This result can explain some experiments by Fainerman \cite{Fainerman} who observed a very weak concentration dependence of the dynamic surface tension for sodium dodecyl sulfate solutions with added salt (Nacl) (the presence of salt makes the behavior of this ionic surfactant similar to neutral ones).

In conclusion, we have presented a theory of surfactant adsorption from a micellar solution
that includes the mechanism of micelles relaxation and which is a generalization of a free-energy formulation for submicellar solution \cite{ad}. Contrary to previous models that assume a bulk diffusion-limited process, we can include the possibility that the adsorption is limited by the kinetics at the interface. Whereas for submicellar solution, common non-ionic surfactants molecules usually undergo a diffusion limited adsorption, we showed that the adsorption of free surfactants from a micellar phase could be either diffusion or kinetically limited, according to the value of the characteristic relaxation time of the micelle. These predictions are in qualitative agreement with experiments.

It is a pleasure to thank J.F. Joanny for many valuable discussions.

\end{document}